\documentclass[english,prl,aps,twocolumn,10pt]{revtex4-2}

\usepackage{amsmath}
\usepackage{amssymb}
\usepackage{graphicx}
\usepackage[english]{babel}
\usepackage{array}
\usepackage{verbatim}
\usepackage[colorlinks=true, pdfstartview=FitV, linkcolor=blue, citecolor=blue, urlcolor=blue]{hyperref} % enable links
\usepackage{oplotsymbl} % special symbols, like \pentago

\newcommand{\ket}[1]{\ensuremath{\left|#1\right\rangle}}

\newcommand{\mr}[1]{\mathrm{#1}}
\newcommand{\unit}[1]{\,\mathrm{#1}}
\newcommand{\um}{\,\mu{\rm m}}
\newcommand{\us}{\,\mu{\rm s}}

\newcommand{\uC}{\,\mu{\rm C}}

\newcommand{\ye}{\gamma_\mr{e}}

\newcommand{\AlO}{Al$_2$O$_3$}
\newcommand{\PZT}{Pb[Zr$_x$Ti$_{1-x}$]O$_3$}
\newcommand{\SrRuO}{SrRuO$_3$}
\newcommand{\SrTiO}{SrTiO$_3$}
\newcommand{\YMnO}{YMnO$_3$}

\newcommand{\vecB}{\bf B}
\newcommand{\vecE}{\bf E}
\newcommand{\vecK}{\bf K}
\newcommand{\vecn}{\bf n}
\newcommand{\vecP}{\bf P}

\newcommand{\Eac}{E_\mr{ac}}
\newcommand{\Eperp}{E_\perp}
\newcommand{\fc}{f}
\newcommand{\kperp}{k_\perp}
\newcommand{\xosc}{x_\mr{osc}}
\newcommand{\xNV}{x_\mr{NV}}
\newcommand{\yNV}{y_\mr{NV}}
\newcommand{\zNV}{z_\mr{NV}}

\newcommand{\ep}{\epsilon_{0}}
\newcommand{\phiB}{\varphi_B}
\newcommand{\phiE}{\varphi_E}
\newcommand{\phidE}{\varphi_{\Eac}}
\newcommand{\phiNV}{\varphi_\mr{NV}}
\newcommand{\phiS}{\varphi_\xi}
\newcommand{\thetaB}{\theta_B}
\newcommand{\thetaNV}{\theta_\mr{NV}}

\begin{document}

\title{Imaging ferroelectric domains with a single-spin scanning quantum sensor}

\author{W.~S.~Huxter$^{1}$, M.~F.~Sarott$^{2}$, M.~Trassin$^{2}$, and C.~L.~Degen$^{1,3}$}

\affiliation{$^1$Department of Physics, ETH Zurich, Otto Stern Weg 1, 8093 Zurich, Switzerland.}
\affiliation{$^2$Department of Materials, ETH Zurich, Vladimir Prelog Weg 1-5/10, 8093 Zurich, Switzerland.}
\affiliation{$^3$Quantum Center, ETH Zurich, 8093 Zurich, Switzerland.}

\email{degenc@ethz.ch}

%\date{\today}

\maketitle

% first paragraph
\textbf{
The ability to sensitively image electric fields is important for understanding many nanoelectronic phenomena, including charge accumulation at surfaces~\cite{yoo97} and interfaces~\cite{schoenherr19} and field distributions in active electronic devices~\cite{cao21}.  A particularly exciting application is the visualization of domain patterns in ferroelectric and nanoferroic materials~\cite{gruverman19,lachheb21} owing to their potential in computing and data storage~\cite{martin16,sharma21,meier21}.
Here, we use a scanning nitrogen-vacancy (NV) microscope, well known for its use in magnetometry~\cite{rondin14}, to image domain patterns in piezoelectric (\PZT) and improper ferroelectric (\YMnO) materials through their electric fields.  Electric field detection is enabled by measuring the Stark shift of the NV spin~\cite{vanoort90,dolde11} using a gradiometric detection scheme~\cite{huxter22}.  
Analysis of the electric field maps allows us to discriminate between different types of surface charge distributions, as well as to reconstruct maps of the three-dimensional electric field vector and charge density.
The ability to measure both stray electric and magnetic fields~\cite{doherty12,rondin14} under ambient conditions opens exciting opportunities for the study of multiferroic and multifunctional materials and devices~\cite{fiebig16,meier21}.
}

%%% Introduction

Real-space imaging of electric fields at the nanoscale is an important aim across many emerging fields, as near-surface fields are tied to the electrical polarization or charge distribution of the underlying system.  Sensitive imaging of nanoscale electric phenomena has been demonstrated by a number of techniques, most prominently by electrostatic force microscopy~\cite{saurenbach90} and piezoresponse force microscopy (PFM)~\cite{gruverman19}, along with the related techniques of Kelvin probe force microscopy~\cite{nonnenmacher91}, low-energy electron microscopy~\cite{lachheb21}, and emerging scanning quantum technologies~\cite{yoo97,wagner15}.  However, most of these techniques are limited to low temperatures or high-vacuum conditions, require thin-film samples and back electrodes, and measure indirect quantities, such as piezoelectric coefficients or surface potentials.  NV centers in diamond~\cite{doherty12,rondin14,taylor08} provide a path to quantitatively image electric fields under ambient conditions, do not require back electrodes or applied voltages, and measure a quantity directly proportional to the surface polarization.

In this work, we apply scanning NV microscopy to map static electric stray fields above surfaces with sub-100\,nm resolution.  Using mechanical oscillation of the tip~\cite{huxter22} to overcome the challenges of static screening~\cite{oberg20} and low coupling to electric fields~\cite{vanoort90,dolde11}, we reach an excellent sensitivity of $0.24\unit{kV\,cm^{-1}\,Hz^{-1/2}}$, on par with the sensitivities demonstrated for ac field detection in bulk diamond~\cite{dolde11}.  We illustrate the impact of our approach by imaging patterned domains in application-relevant ferroelectric thin films and by mapping the natural domain configuration in a prototypical improper ferroelectric.

Electric field sensing with NV centers relies on a local electric Stark effect.  The Stark effect causes a shift in the NV spin energy levels that is measured using optically-detected magnetic resonance (ODMR) \cite{vanoort90,dolde11}.  The Stark effect is anisotropic and largest in the transverse plane of the NV center, which sits perpendicular to the anisotropy axis ($\zNV$ axis). This leads to an in-plane electric field coupling that is approximately $50\times$ larger than the out-of-plane coupling \cite{dolde11,doherty12}.
To maximize the in-plane electric field response and simultaneously suppress the response to magnetic fields, a small magnetic bias field is applied transverse to the NV anisotropy axis.  Electric field detection in this configuration has previously been demonstrated in bulk diamond, where electric fields are created with external electrodes~\cite{dolde11,iwasaki17,michl19}, charged scanning probes~\cite{barson21,bian21}, surface band bending~\cite{broadway18ne}, and intrinsic dopant charges~\cite{dolde14}.

%%% Experimental

\begin{figure}
	\includegraphics[width=1.0\columnwidth]{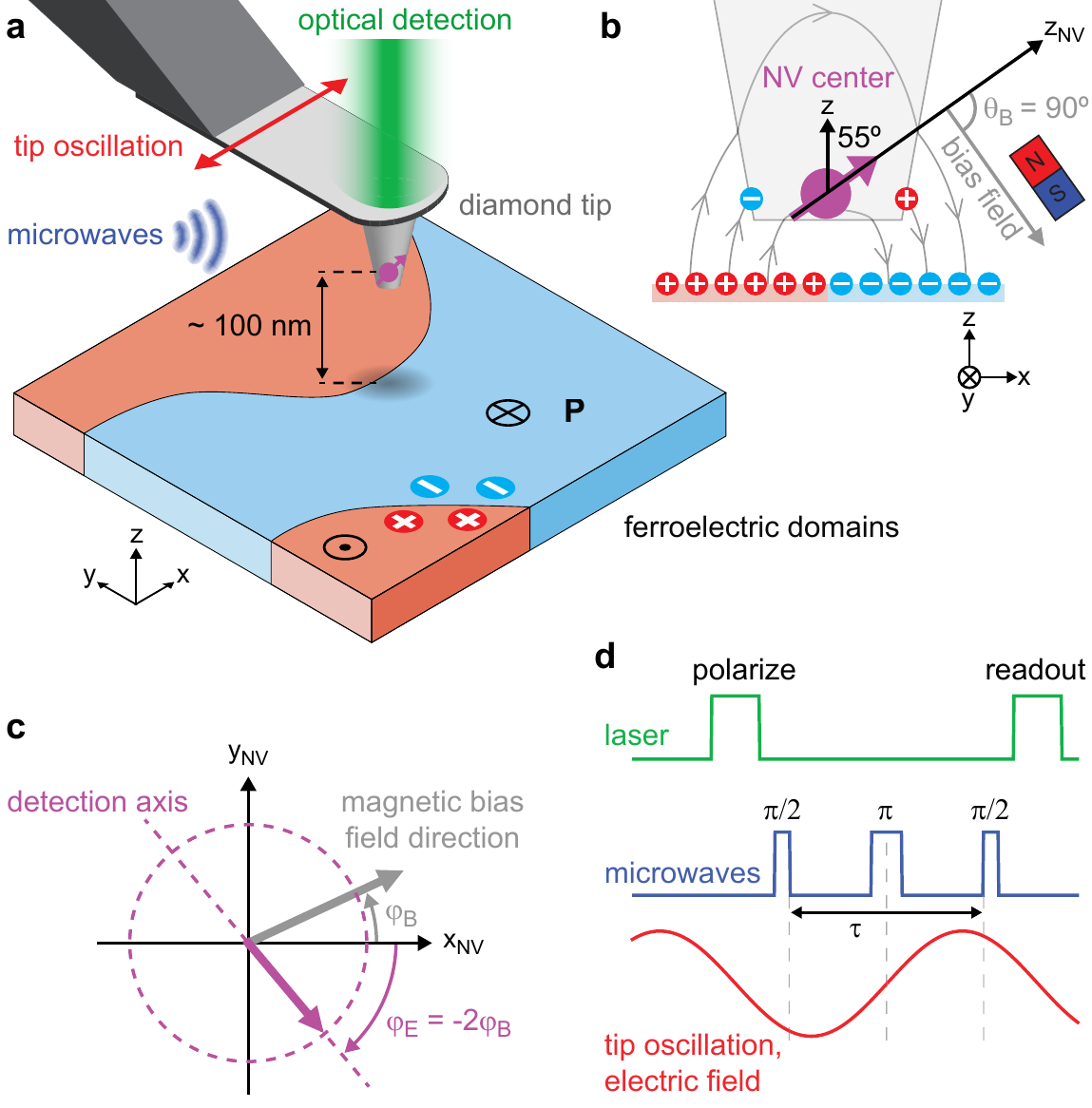}
	\caption{
		{\bf Scanning NV electrometer.}
		{\bf a}, The NV center is located at the apex of the diamond tip and is oscillated in shear-mode through the electrical drive of a tuning fork (not shown) while it is scanned over the surface.  Laser and microwave pulses synchronized to the drive are used for signal readout.  A three-axis piezo stage underneath the sample is used for positioning.
		{\bf b}, Geometry of tip and sample with the vector orientations of the NV spin, magnetic bias field and sample electric field.  Sample surface charges and screening charges on the tip are also shown. 
		{\bf c}, The in-plane electric field detection axis (purple) is determined by the in-plane direction of the magnetic bias field (gray). ($x,y,z$) and ($\xNV,\yNV,\zNV$) denote the laboratory and NV frames of reference (see Methods for definition).
		{\bf d}, Spin-echo pulse scheme synchronized to the tip oscillation for the ac measurement of the electric field gradient.
	}
	\label{fig1}
\end{figure}

In our experiment, the NV center is embedded in the tip of a diamond scanning probe (Fig.~\ref{fig1}a).
The scanning probe arrangement allows us to extend electric field sensing to image general materials systems, including ferroelectrics.  We mount the diamond probe on a quartz tuning fork oscillator providing force-feedback for safe approach and scanning. Owing to the tip fabrication procedure, the NV anisotropy axis is $\sim 35^\circ$ away from the scan plane (Fig.~\ref{fig1}b).

To enable $E$-field detection, we accurately orient an external bias field of $5-12\unit{mT}$ transverse to $\zNV$ ($\thetaB = 90^\circ$) with $\sim 0.5^\circ$ of uncertainty (Methods).  In this bias field configuration, the spin transition frequencies $\omega_\pm$ are linearly sensitive to electric fields,
\begin{align}
\omega_\pm \approx \omega_\pm^{(0)} \mp 2\pi\kperp\Eperp\cos(2\phiB+\phiE)
\label{eq:omega}
\end{align}
while correcting for magnetic fields up to second order \cite{dolde11,doherty12}.  Here, $\omega_\pm^{(0)}$ are the spin resonance frequencies in absence of the electric field (Supplementary Section 1), $\kperp = 16.5\unit{Hz\,V^{-1}\,cm}$ is the coupling constant \cite{michl19}, $\Eperp$ is the magnitude of the in-plane $E$-field vector, and $\phiB = \arctan(B_{\yNV}/B_{\xNV})$ and $\phiE = \arctan (E_{\yNV}/E_{\xNV})$ are the in-plane angles of the magnetic bias field and electric field vectors, respectively (Fig.~\ref{fig1}c).  Eq.~(\ref{eq:omega}) neglects strain interactions (Methods).  The angular dependence on the bias field results in a maximal frequency shift when $\phiE = -2\phiB$ (modulo $\pi$), defining the detection axis (Fig.~\ref{fig1}c).  To polarize, manipulate and detect the NV spin state we use a combination of laser and microwave pulses together with a single-photon counting module (Fig.~\ref{fig1}a,d and Methods). 

\begin{figure*}
	\includegraphics[width=1.0\textwidth]{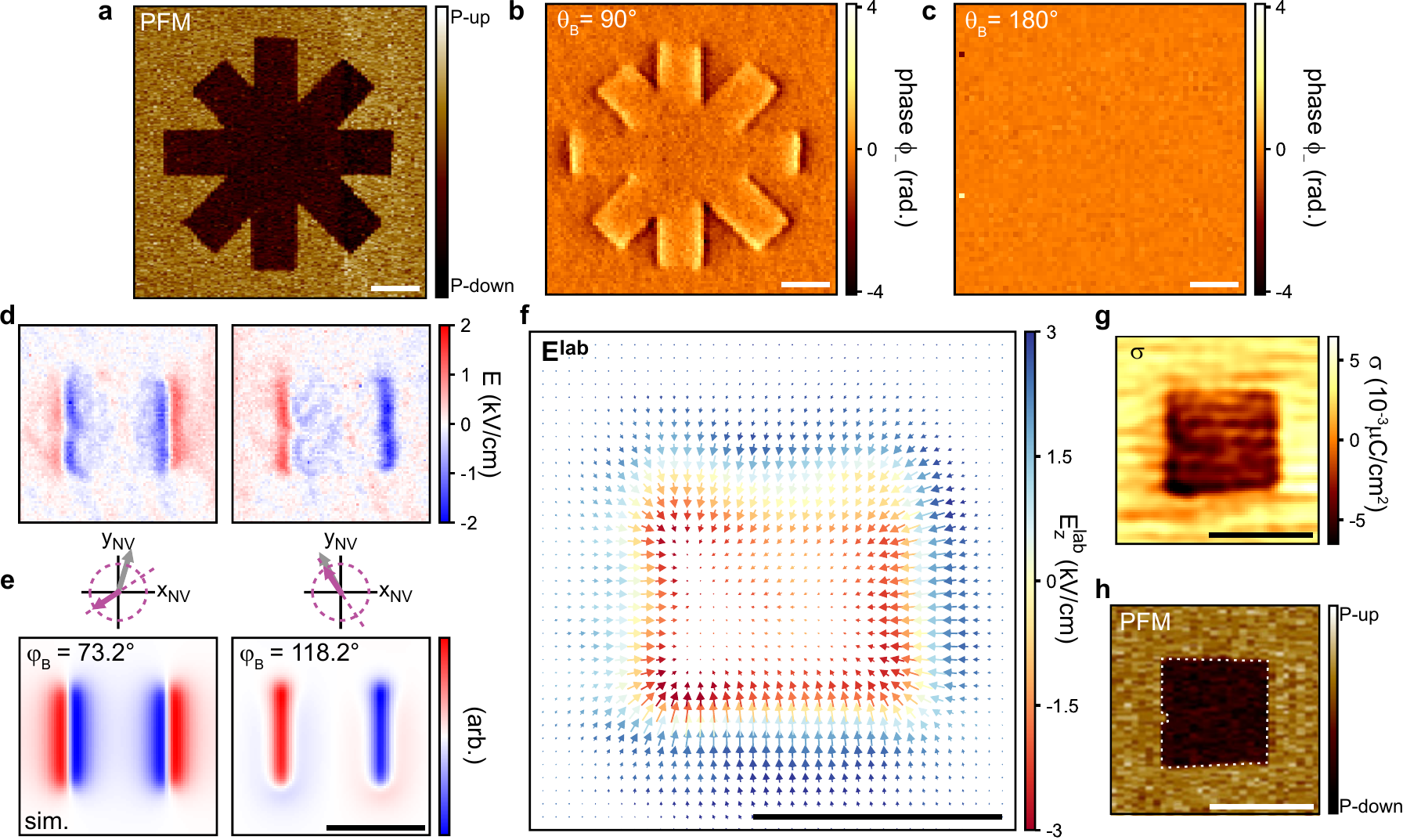}
	\caption{
		{\bf Vector electrometry of a piezoelectric PZT film.}
		{\bf a}, PFM image of a patterned eight-crossed domain structure.
		{\bf b}, NV electrometry taken over the same region with the bias magnet oriented in-plane ($\vecB \perp \zNV$).
		{\bf c}, Same as \textbf{b} but with the bias magnet oriented out-of-plane ($\vecB || \zNV$).
		{\bf d}, NV electrometry maps from a patterned square domain. The two maps show orthogonal in-plane field components, obtained by shifting the magnetic angle $\phiB$ by $45^\circ$.
		{\bf e}, Corresponding simulated electric field images with schematics of the magnetic bias angle (gray) and detection axis (purple).
		{\bf f}, Laboratory-frame vector plot of the electric field ${\vecE}^\mr{lab}=(E_x^\mr{lab},E_y^\mr{lab},E_z^\mr{lab})$ reconstructed from the maps in panel \textbf{d}. $E_x^\mr{lab}$ and $E_y^\mr{lab}$ components are represented as arrows and $E_z^\mr{lab}$ is shown as a color.
		{\bf g}, Reconstructed surface charge density $\sigma$ revealing the written square domain pattern.
		{\bf h}, Corresponding PFM image.
		Dwell times are 12 s per pixel ({\bf b, c}) and 10 s per pixel ({\bf d}).
		Scale bars, 1$\unit{\um}$.
	}
	\label{fig2}
\end{figure*}

An important concern with static field measurements is screening of electric fields by mobile charges on the diamond tip (Fig.~\ref{fig1}b).  This issue has hindered previous attempts at implementing a scanning NV electrometer~\cite{oberg20,lorenzelli21}.  To overcome this screening, we oscillate the diamond sensor using the mechanical resonance ($\fc \sim 32\unit{kHz}$) of the tuning fork and detect the resulting ac electric field (Fig.~\ref{fig1}d).  This ac electric signal is proportional to the electric field gradient in the oscillation direction \cite{huxter22}.  In addition to alleviating static screening, the ac detection also improves sensitivity by at least an order of magnitude~\cite{taylor08,dolde11,huxter22}.  For the spin-echo pulse sequence shown in Fig.~\ref{fig1}d, the field-induced coherent phase accumulation of the NV spin is
$\phi_\pm = \pm 4\kperp \Eac \cos(2\phiB+\phidE) \sin^2(\pi \fc \tau/2)/\fc$
where $\Eac = \xosc \sqrt{(\partial_x E_{\xNV})^2 + (\partial_x E_{\yNV})^2}$ and $\phidE = \arctan(\partial_x E_{\yNV}/\partial_x E_{\xNV})$, $\xosc$ is the oscillation amplitude along $x$, and $\tau$ is the evolution time (Supplementary Section 1).

%%% Results

We demonstrate scanning electrometry by imaging the electric fields appearing above the surfaces of two ferroelectric materials.  Our first sample is a $50\unit{nm}$-thick film of out-of-plane polarized lead zirconate titanate (\PZT, PZT) grown on top of a \SrRuO-buffered (001)-oriented \SrTiO\ substrate.  PZT, the most technologically important ferroelectric, has a large polarization ($|{\vecP}| \sim 75\unit{\uC\,cm^{-2}}$) ideal for an initial demonstration.  To create recognizable structures, we write a series of ferroelectric domain patterns by locally inverting the polarization of the film using a conductive atomic force microscopy (AFM) tip.  Fig.~\ref{fig2}a shows an image of the out-of-plane PFM contrast above one of these patterns.

Fig.~\ref{fig2}b presents an NV electrometry map taken above the same location. Owing to our gradiometric detection scheme, the signal is maximum near vertical edges of the pattern. This directionality reflects the horizontal oscillation direction of the sensor and other oscillation directions (such as a tapping mode) may be used to acquire different spatial signatures \cite{huxter22}. To verify that the signal is indeed due to electric fields, we purposely misalign the bias field to $\thetaB=95^\circ$ (Fig.~S2) and $180^\circ$ (Fig.~\ref{fig2}c).  As expected, the signal disappears under the bias field misalignment as the NV center becomes insensitive to electric fields.  We additionally tried detecting the $E$-field in a dc sensing mode and observed no signal (Fig.~S3).  Thus, dynamic (ac) operation is essential for overcoming screening and enabling static electric field sensing.  Detection at higher frequencies, and with multi-pulse measurement schemes is shown in Fig.~S1 and S3. 

By rotating the in-plane angle $\phiB$ of the bias field, we can rotate the in-plane detection axis (Fig.~\ref{fig1}c)~\cite{dolde11}. In particular, by acquiring electric field maps shifted in $\phiB$ by $45^\circ$, it becomes possible to map two orthogonal components of the $E$-field signal.  Fig.~\ref{fig2}d shows such orthogonal electric gradient maps from a square domain. The experimental maps are in good agreement with  numerical simulations of a square domain with constant surface charge (Fig.~\ref{fig2}e).  By combining the orthogonal field components we are able to reconstruct the full three-dimensional electric field vector above the domain (Fig.~\ref{fig2}f and Methods).
In principle, field maps like Fig.~\ref{fig2}f could allow measurement of the domain wall width and possibly the chirality~\cite{tetienne15}. 
However, our spatial resolution ($\sim 100\unit{nm}$,~Fig.~S6) is not yet sufficient to resolve the structure of the $<10\unit{nm}$ domain walls in PZT~\cite{deluca16}.
Using reverse propagation of Coulomb's law, we also compute the equivalent surface charge density $\sigma$ at the top surface of the ferroelectric film (Fig.~\ref{fig2}g and Methods), where $\sigma = \vecP\cdot\vecn$ and $\vecn$ is the surface normal. The reconstruction involves two charge sheets of opposite sign since further analysis (below) reveals the presence of charges beneath the surface. Our result is in excellent agreement with the out-of-plane PFM image shown in Fig.~\ref{fig2}h, where both a defect and the asymmetric shape of the domain are reproduced.

\begin{figure}
    \includegraphics[width=1.0\columnwidth]{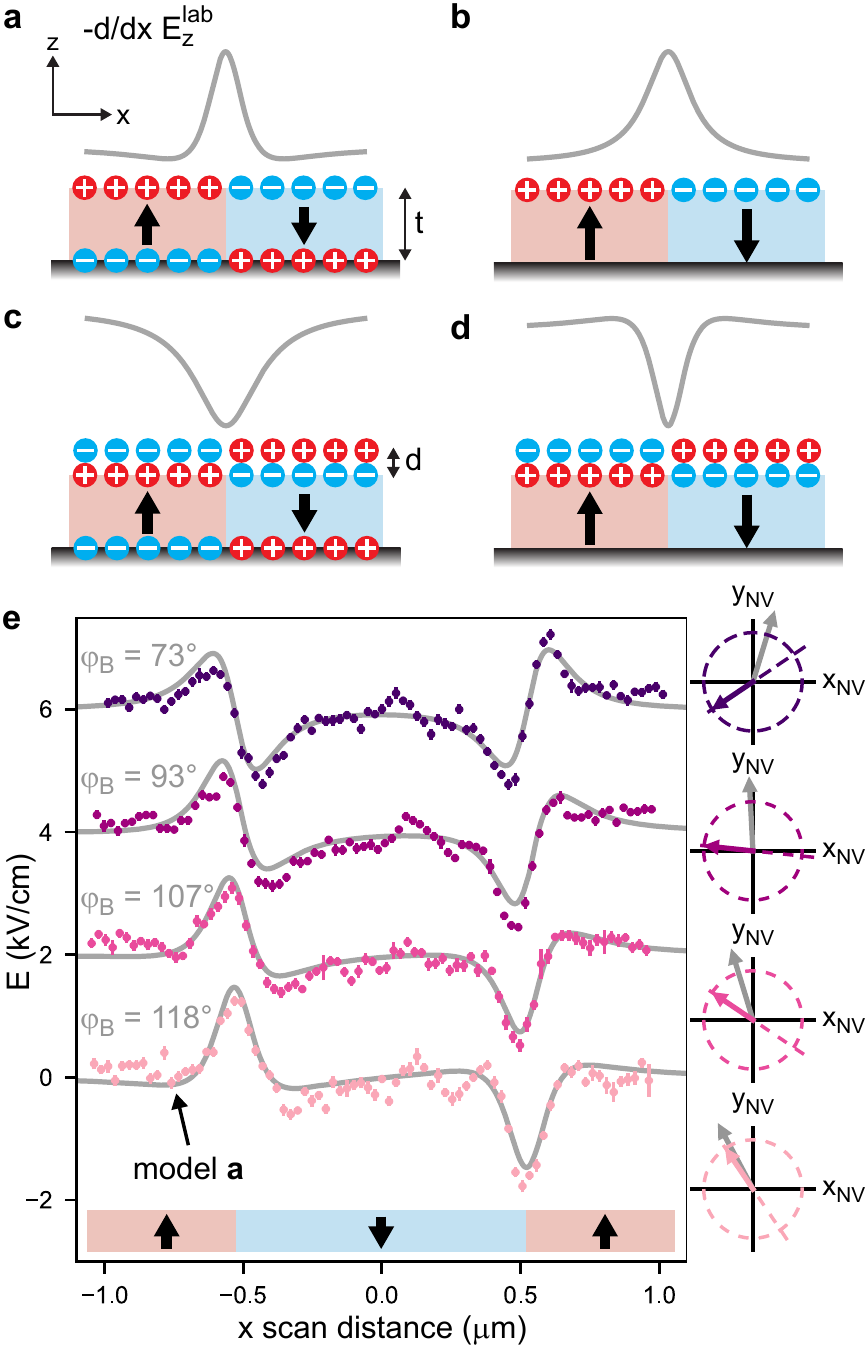}
    \caption{{\bf Surface charge models at domain walls in PZT.}
				{\bf a-d}, Surface charge models and negative $z$ component of the laboratory electric field gradient ($\partial_x E_z^\mr{lab}$) over a domain wall, which corresponds to the NV signal for $\phiB = 121^\circ$.
				{\bf a}, Ideal bound charge model with monopole charge sheets on the top and bottom surfaces. $t$ is the sample thickness and for thin films the field gradient is dipole-like.
				{\bf b}, Same as {\bf a} with complete screening from the bottom electrode.
				{\bf c}, Same as {\bf a} with an adsorbed top layer separated by a distance $d$. The adsorbed layer screens the top surface and the signal is mainly produced by the bottom layer.
				{\bf d}, Same as {\bf c} with complete screening from the bottom electrode, resulting in a dipole surface.
				{\bf e}, NV electrometry line scans taken across the square domain shown in Fig.~\ref*{fig2}d.  Bias field angles $\phiB$ are listed on the left and schematically shown (with detection axis) on the right.  Profile fits (gray) are only compatible with surface charge model~{\bf a}.  Error bars are shot-noise propagated uncertainties. 
		}
    \label{fig3}
\end{figure}

To further interpret our measurements we develop simple models of surface charge distributions and their associated electric field gradients, Figs.~\ref{fig3}(a-d). By comparing these to the experimental line scans (Fig.~\ref{fig3}e), we can discriminate between different surface charge scenarios and extract quantitative information on the surface charge density $\sigma$.
We find that our data are best fit by the model shown in Fig.~\ref{fig3}a, which includes two layers of opposite charges on the top and bottom of the PZT sample. The other models (Fig.~\ref{fig3}b-d) are inconsistent with the polarity or shape of the experimental line scans.
In particular, the Lorentzian shape of the $E_z^\mr{lab}$ field gradient produced by a net surface charge (models b and c) fails to reproduce the dips in the signal at either side of the domain.  Additionally, models c and d would result in a polarity opposite to what is expected from the known polarization of our PZT sample.
The ability to distinguish between different charge models and quantify the screening efficiency of a back electrode will be useful for analyzing charge dynamics and screening behavior at ferroelectric domain walls, interfaces and surfaces~\cite{kalinin18}, even once the materials are buried in a device architecture~\cite{strkalj19}.

\begin{figure*}
	\includegraphics[width=0.65\textwidth]{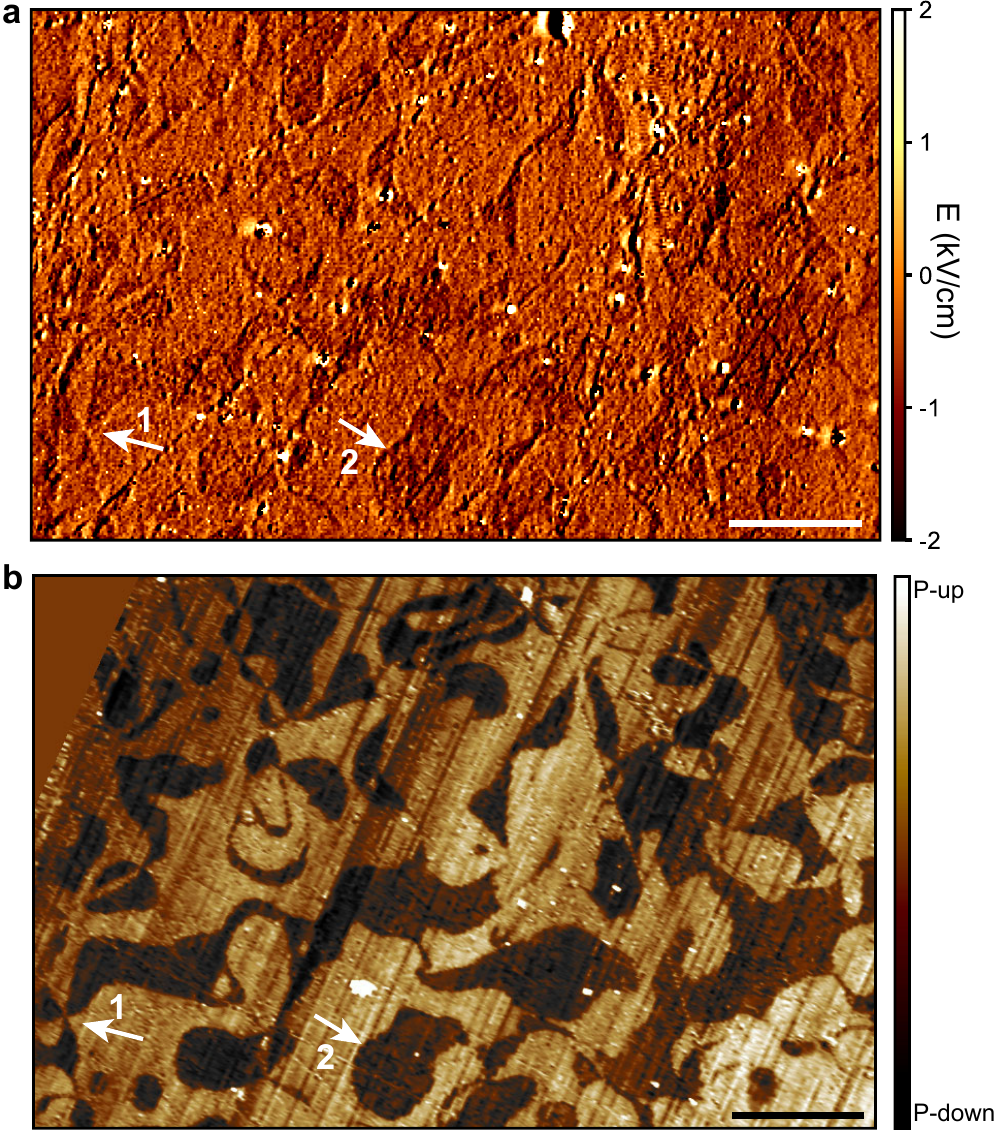}
	\caption{
	  {\bf Naturally occurring ferroelectric domain pattern in hexagonal \YMnO.}
		 {\bf a}, NV electrometry on \YMnO. Domains appear with constant contrast, which may be attributed to tip oscillations that deviate from pure shear mode (Supplementary Fig. S3).  Bright features reflect patch charges at topographic defects.
		 {\bf b}, Out-of-plane PFM on \YMnO\ over the same region. Artifacts from topographic cross-talk are not observed in panel {\bf a}. 
		 White arrows indicate (1) a vortex domain and (2) a 180$^\circ$ domain wall.
		 Scale bar, $5\unit{\um}$.
	}
	\label{fig4}
\end{figure*}

Fits to the line scans in Fig.~\ref{fig3}e yield an effective surface charge density of approximately $\sigma = (3.5\pm0.3)\times 10^{-3} \unit{\uC\,cm^{-2}}$. The fits use the known bias field direction, NV center orientation and stand-off distance (Methods and Supplementary Section 2).
This surface charge density is about four orders of magnitude lower than the expected value for our PZT sample ($75 \unit{\uC\,cm^{-2}}$, Ref.~\cite{suleiman21}). The large difference may be attributed to a number of screening mechanisms on both the sample and diamond tip.
This includes surface screening from adsorbates~\cite{segura13} or the formation of a charged, off-stoichiometric surface layer~\cite{strkalj20} at the surface of the film. 
On the diamond tip, screening could be a result of the dielectric constant ($\epsilon_r = 5.7$) as well as partial screening from an adsorbed water layer and mobiles charges still present at $\sim 32 \unit{kHz}$ \cite{oberg20}.  Calibration against a known electrode may allow disentangling tip and sample screening~\cite{qiu22}.

%%% Results on YMnO3

We further illustrate scanning electrometry by imaging the natural domain pattern of an improper ferroelectric, the hexagonal manganite \YMnO.  Hexagonal manganites are exciting benchmark materials because their surface polarization ($|{\vecP}| \sim 5.5\unit{\uC\,cm^{-2}}$ for \YMnO) is over an order of magnitude smaller than our PZT sample and typical for ferroelectric and multifunctional materials~\cite{fiebig16,meier21}.  In addition, hexagonal manganites are type-I multiferroics and become antiferromagnetically ordered below $T_N \sim 100\unit{K}$~\cite{lilienblum15,fiebig16}.  While the antiferromagnetic domain pattern has been imaged with scanning NV magnetometry at cryogenic temperatures~\cite{lorenzelli21}, here, we focus on the ferroelectric domain pattern accessible under ambient conditions.
  
Fig.~\ref{fig4}a shows an electrometry map recorded above a polished, bulk \YMnO\ sample.  Although the signal-to-noise ratio is lower compared to the PZT data, as expected from the smaller polarization,  domains (including vortex domains) are clearly visible and resemble the pattern observed by PFM (Fig.~\ref{fig4}b). The NV electrometry image also reveals long, straight line-like features and defects which we interpret as charge accumulation near topographic features such as polishing marks. Correlative magnetic measurements over the same region show no magnetic signal (Fig.~S4).
%%% Conclusion

In summary, we demonstrate nanoscale electric field imaging using a scanning NV microscope.  Crucial to our experiment is the use of gradiometric detection~\cite{huxter22}, which both alleviates static field screening at the diamond surface and greatly improves the sensitivity compared to static sensing schemes.
The electric field sensitivity of $\sim 0.24\unit{kV\,cm^{-1}\,Hz^{-1/2}}$ demonstrated in our work (Methods) is similar to those shown in the bulk~\cite{dolde11}.

By imaging ferroelectric domains through their stray electric fields we demonstrate a complementary imaging method to existing scanning probe techniques.
While the spatial resolution ($\sim 100\unit{nm}$) demonstrated in our work is currently outmatched by these techniques~\cite{gruverman19,saurenbach90,nonnenmacher91}, moving the tip closer to the sample surface is expected to lower this value to below 50\,nm, and perhaps below 30\,nm~\cite{ariyaratne18}.
Moreover, the long measurement times of several seconds per pixel can be reduced through increased dynamical decoupling and excited-state spectroscopy at cryogenic temperatures~\cite{block21}.  The benefits of NV electrometry lie in it's non-perturbing aspect, as it can quantitatively measure an unknown electric dipole configuration (electrical polarization and buried charged planes) without the need of back electrodes or applied voltages, all without topographic crosstalk in the measurement signal.

Looking towards future applications, an intriguing perspective is the correlative imaging of magnetic and electric fields in multiferroic and multifunctional materials~\cite{lilienblum15,chauleau20}. Since the sensitivity to electric and magnetic fields can be adjusted by a simple re-orientation of the magnetic bias field, magnetic and electric field maps can be recorded from the same sample region without breaking experimental conditions. This will allow analyzing multiple order parameters \textit{in situ}.  Together, the multimodal capability opens exciting opportunities to study magneto-electric coupling and detail the formation and structure of domains and domain walls in these multifaceted and technologically relevant materials.
Lastly, the ability to probe buried charged surfaces may be instrumental in the pursuit of energy-efficient device concepts based on electric field control of magnetization~\cite{manipatruni19,noel20}.

%%%%%%%%%%%%%% Note

\textit{Note added:}
While finalizing this manuscript, we became aware of related work~\cite{qiu22} describing scanning NV electrometry using a similar gradiometry technique to image electric fields from electrodes.

%%%%%%%%%%%%%% Acknowledgments

\vspace{0.5cm}\textbf{Acknowledgments}

The authors thank M.L. Palm and P.J. Scheidegger for helpful comments on the manuscript, M.W. Doherty, S. Ernst, N. Prumbaum and P. Welter for useful discussions, L.M. Giraldo and M. Fiebig for providing the \YMnO\ sample, and QZabre AG for support in probe fabrication.
This work was supported by the European Research Council through ERC CoG 817720 (IMAGINE), the Swiss National Science Foundation (SNSF) through the NCCR QSIT, a National Centre of Competence in Research in Quantum Science and Technology, Grant No. 51NF40-185902, and the Advancing Science and TEchnology thRough dIamond Quantum Sensing (ASTERIQS) program, Grant No. 820394, of the European Commission.
M.T. acknowledges the SNSF through Project Grant No. 200021-188414, and 
M.T. and M.F.S. acknowledge the SNSF through Spark Grant CRSK-2\_196061.

\vspace{0.5cm}\textbf{Author contributions}

C.L.D. and W.S.H. conceived the experiment.
W.S.H. carried out scanning NV experiments and performed the data analysis.
M.F.S. and M.T. fabricated the PZT sample, wrote the domains and carried out PFM experiments on PZT and \YMnO.
W.S.H. and C.L.D. wrote the manuscript.
M.F.S. and M.T. provided insights on electrically ordered materials.
All authors discussed the results.

%%%%%%%%%%%%% Statements

\vspace{0.5cm}\textbf{Competing interests}

The authors declare no competing interests.

\vspace{0.5cm}\textbf{Data availability statement}

The data that support the findings of this study are available from the corresponding author upon reasonable request.

\vspace{0.5cm}\textbf{Additional information}

Supplementary information accompanies this paper. Correspondence and requests for materials should be addressed to C.L.D.

%%%%%%%%%%%%% Methods

\vspace{0.5cm}\textbf{Methods}\vspace{0.5cm}

{\bf Experimental set-up}

Experiments are performed at room temperature with a custom-built scanning NV microscope. Micro-positioning is carried out by a closed-loop three-axis piezo stage (Physik Instrumente) and AFM feedback control is carried out by a lock-in amplifier (HF2LI, Zurich Instruments). Photoluminescence (PL) of the NV centers is measured with an avalanche photodiode (Excelitas) and data are collected by a data acquisition card (PCIe-6353, National Instruments). Microwave pulses and sequences are created with a signal generator (Quicksyn FSW-0020, National Instruments) and modulated with an IQ mixer (Marki) and an arbitrary waveform generator (HDAWG, Zurich Instruments). NV centers are illuminated at $<100\unit{\mu W}$ by a custom-designed $520\unit{nm}$ pulsed diode laser. Scanning NV tips are purchased from QZabre AG. Two tips are used throughout this study.  Tip \#1 (Figs.~\ref{fig2} and \ref{fig3}) had a stand-off distance of $z=95 \pm 1\unit{nm}$ and tip \#2 (Fig.~\ref{fig4}) had a stand-off distance of $z=61 \pm 2\unit{nm}$ (not including the $20~\unit{nm}$ retract distance used while imaging), determined with a magnetic stripe~\cite{hingant15}.  Oscillation amplitudes  were $\sim 46 \unit{nm}$ in Figs.~\ref{fig2}b,d and~\ref{fig3}e,  $\sim 92\unit{nm}$ in Fig.~\ref{fig2}c, and $\sim 52\unit{nm}$ in Fig.~\ref{fig4}a, determined via stroboscopic imaging~\cite{huxter22}. A movable permanent neodymium magnet (supermagnete) below the sample served as the bias field source. Field alignment is possible by applying a fitting algorithm, which used a numerical model of the field produced by the magnet and NV resonance frequencies as a function of magnet position.  The drift stability of the microscope is $<30\unit{nm}$ per day and no drift correction techniques are employed.

{\bf Spin energy levels in a transverse magnetic field}

With an off-axis field, the usual spin-state description (using $m_\mr{s}$) of the NV center is not ideal, as the magnetic quantum number $m_\mr{s}$ is no longer conserved.  In this scenario the eigenstates, which we denote as $\ket{0}$, $\ket{-}$, and $\ket{+}$, are superpositions of the usual $\ket{m_\mr{s}}$ states~\cite{doherty12,dolde14thesis}. Spin state mixing results in a reduced PL and optical contrast which worsened the overall measurement sensitivity. This effect, however, is manageable for relatively weak ($< 12 \unit{mT}$) bias fields \cite{tetienne12}. Bias fields $< 5 \unit{mT}$ are also non-ideal as $^{15}$N hyperfine coupling effectively misaligned small bias fields.

{\bf Alignment of the transverse magnetic field}

The alignment algorithm of the bias field provided the ability to align with roughly $1^\circ$ of uncertainty, however we improved the alignment by additionally considering hyperfine interactions~\cite{dolde14thesis}. In an off-axis bias field ($\theta_B = 90^\circ$) the $\ket{0}$ state splits into two states that differ by $\Delta = 2 a_{\mr{hf}}\ye B/D$, where $a_{\mr{hf}} = 3.65 \unit{MHz}$ is the off-axis $^{15}$N hyperfine coupling parameter, $\ye = 2\pi\times28\unit{GHz/T}$ is the gyromagnetic ratio of the electron, $B$ is the applied magnetic field and $D=2.87 \unit{GHz}$ is the zero-field-splitting~\cite{dolde14}. When deviating from $\theta_B = 90^\circ$ the on-axis component splits each of the the $\ket{-}$  and $\ket{+}$ states through $^{15}$N hyperfine interaction resulting in eight total transitions (four for $\omega_-$ and four for $\omega_+$). We swept the fitted polar angle a few degrees around $90^\circ$ (with the same field magnitude) and tracked the $\omega_-$ hyperfine resonances. The best alignment is achieved when only two resonances are visible and when the resonance frequency is the largest (on-axis contributions decreased the resonance frequency).

{\bf Influence of crystal strain}

Internal strain in the diamond acts equivalently as a permanent dc $E$-field via the piezocoupling coefficient~\cite{maze11}. For in-plane strains much weaker than electric signals (which is assumed during our analysis) the effect of strain is unimportant. For large in-plane strains (relative to the electric signal), it is still possible to carry out scanning gradiometry without a loss in sensitivity. In this case, the detection axis is controlled by the strain direction (Supplementary Section 1). The angular dependence changes from $\cos \left( 2 \phiB + \phidE \right)$ for small strains to  $\cos(\phiS - \phidE) \cos \left( 2 \phiB + \phiS \right)$, where  $\phiS$ is the in-plane strain angle.

{\bf Laboratory and NV center frames of reference}

To translate between the laboratory frame and NV center frame a representation of the NV center's crystallographic coordinate system is determined in the laboratory frame. 
The laboratory frame is defined by the vectors $\hat{x} = [1, 0, 0]$, $\hat{y} = [0, 1, 0]$, and $\hat{z} = [0, 0, 1]$. The unit vector along the symmetry axis of the NV center ($\hat{z}_\mr{NV})$, pointing from the nitrogen atom towards the vacancy site, is chosen as the $[111]$ crystallographic direction.
The $x$ unit vector ($\hat{x}_\mr{NV}$) is taken to be orthogonal to $\hat{z}_\mr{NV}$ and pointing from the vacancy site towards one of the three nearest carbon atoms ($[11\overline{2}]$ for example, although there are three possible choices)~\cite{doherty12,doherty14}.
Then $\hat{y}_\mr{NV} = \hat{z}_\mr{NV} \times \hat{x}_\mr{NV}$ to preserve right-handedness. To translate between the crystallographic frame and laboratory frame the bias field alignment algorithm and known (001) cut of the diamond tip is used. For example, with the NV center angles of $\thetaNV = 55^\circ$ and $\phiNV = 0^\circ$ (as shown in Fig.~\ref{fig1}b), we get $\hat{x}_\mr{NV} = \frac{1}{\sqrt{3}}[1, 0, -\sqrt{2}]$, $\hat{y}_\mr{NV} = [0, 1, 0]$, and $\hat{z}_\mr{NV} = \frac{1}{\sqrt{3}}[\sqrt{2}, 0, 1]$. It is important to note that using a different NV center reference frame definition can result in an incorrect computation of laboratory frame electric fields (Supplemental Section 5).

{\bf Gradiometry technique}

A complete description of scanning gradiometry, including calibration procedures, can be found in Ref.~\onlinecite{huxter22}.  A $\sim 2\unit{\us}$ laser pulse is used to polarize the NV center into the $m_\mr{s} = 0$ state, which for small off-axis bias fields corresponds to the $\ket{0}$ state.  Next, a microwave $\pi/2$-pulse is applied to create a superposition between $\ket{0}$ and one of $\ket{\pm}$. The quantum phase $\phi$ accumulated between the two states during the coherent precession is
$\phi_\pm =  \int_0^\tau g(t) \Delta \omega_\pm (t) dt$,
where $g(t)$ is the modulation function~\cite{degen17},
$\Delta\omega_\pm (t) = \mp 2\pi k_\perp \Eac(t) \cos(2\phiB + \phidE) \sin(2\pi\fc t)$ is the detuning
(see Supplementary Section 1) and $\tau$ is the evolution time.  We used a four-phase cycling technique~\cite{palm22,huxter22} of the last $\pi/2$-pulse to measure $\phi_\pm$.  The readout of the NV center's spin state is performed by another $\sim 2\unit{\us}$ laser pulse, during which the photons emitted from the NV center are collected across a $\sim 600 \unit{ns}$ window.

{\bf Samples}

\noindent\textit{Lead Zirconate Titanate:}
The 50 nm thick \PZT\ film and the 10 nm thick \SrRuO\ electrode are grown on (001)-oriented \SrTiO\ (CrysTec GmbH) using pulsed layer deposition with a KrF excimer laser at 248 nm (LPXpro, Coherent Ltd.). \SrRuO\ is grown at a substrate temperature of 700$^\circ$C with an O$_2$ partial pressure of $0.1 \unit{mbar}$ and a laser fluence of $0.95\unit{Jcm^{-2}}$ at $4\unit{Hz}$. \PZT\ is grown at 550$^\circ$C at $0.12\unit{mbar}$ O$_2$ partial pressure and a laser fluence of $1.2\unit{Jcm^{-2}}$ at $4\unit{Hz}$. The film is subsequently cooled to room temperature under growth pressure. Layer thicknesses are measured using X-ray reflectivity with a four-cycle thin-film diffratometer (PANalytical X'Pert$^3$ MRD, Cu K$\alpha_1$). Topography and PFM experiments are performed on a Bruker Multimode 8 atomic force microscope using Pt-coated Si tips (MikroMasch, $k=5.4 \unit{Nm^{-1}}$).

\noindent\textit{Hexagonal Yttrium Manganite:}
The \YMnO\ bulk crystal is grown by the floating-zone technique, pre-oriented using Laue diffraction and cut perpendicular to the crystal $z$-axis with a diamond saw. The sample is flattened by lapping with a \AlO\ powder in water solution ($9\unit{\um}$ particle size). Subsequently, the sample is chemo-mechanically polished using a colloidal silica slurry. 
To generate domains, the sample is pre-annealed and cooled through $T_C$ in an O$_2$ atmosphere \cite{griffin12}.

{\bf Electric field vector reconstruction}

To reconstruct the $E$-field vector (Fig.~\ref{fig2}f and Fig.~S6) two images recorded with a $45^\circ$ difference in $\phiB$, denoted as $I_1 (x,y) = \Eac\cos(2\phiB + \phidE)$ and $I_2(x,y) = \Eac\sin(2\phiB + \phidE)$, are combined to yield the magnitude ($|\Eac|(x,y)$) and angle ($\phidE(x,y)$) of the measured electric field signal
\begin{align}
	|\Eac|(x,y) &= \sqrt{(I_1(x,y))^2 + (I_2(x,y))^2} \ , \\
	\phidE(x,y) &= \arctan \left( \frac{I_2(x,y)}{I_1(x,y)} \right) - 2\phiB ~ \text{(mod $2\pi$)} \ .
\end{align}
Since there are three possible choices for $\hat{x}_\mr{NV}$, owing to the $C_{3\nu}$ symmetry of the NV center, $\phiB$ and $\phidE(x,y)$ are only known up to a multiple of $2\pi/3$. This propagates to the $E$-field gradients along the $x$-and $y$-directions of the NV center, which are calculated as
\begin{align}
	\xosc \partial_r E_{\xNV}(x,y) &= |\Eac|(x,y)\cos(\phidE(x,y)) \ , \\
	\xosc \partial_r E_{\yNV}(x,y) &= |\Eac|(x,y)\sin(\phidE(x,y)) \ ,
\end{align}
where $\xosc$ is the tip oscillation amplitude and $\partial_r$ is the directional derivative along the unit vector $\hat{r} = \hat{x}\cos\alpha + \hat{y}\sin\alpha$ in the laboratory frame and $\alpha$ is the in-plane oscillation angle. With either of these images it is possible to reconstruct the laboratory components of the $E$-field gradients in Fourier space. For example, with $\hat{x}_\mr{NV}$ the laboratory frame $E$-field gradient vector is
\begin{equation}
	\mathcal{F}\{\partial_r {\vecE}^\mr{lab} \} =  \mathcal{F}\{\partial_r E_{\xNV} \} \frac{\vecK}{\hat{x}_\mr{NV} \cdot {\vecK}} \ ,
\end{equation}
where $\mathcal{F}$ is the Fourier transform, ${\vecK} = [ik_x, ik_y, K]$ and $K = \sqrt{k_x^2 + k_y^2}$. The laboratory frame vector components can be determined independent of the three possible $\hat{x}_\mr{NV}$ and $\hat{y}_\mr{NV}$ choices because the term in the denominator removes its influence. The last step is integration in Fourier space with a wavevector-dependent window function that cuts off high-frequency terms (using a Hann window filter)~\cite{palm22} and a line-filter that removes the amplified noise perpendicular to the direction of integration~\cite{huxter22}. The $E$-field vector is computed with
\begin{equation}
		\mathcal{F}\{{\vecE}^\mr{lab}\} = \frac{\mathcal{F} \{ \partial_r {\vecE}^\mr{lab} \}  W(\lambda, \alpha)}{-i \xosc k_r } \ ,
\end{equation}
where, $k_r = k_x \cos(\alpha) + k_y \sin (\alpha)$ and $W(\lambda, \alpha)$ is the window function. We set the cut-off wavelength to the stand-off distance ($\lambda = z$, which produces a cut-off wavevector of $k = 2\pi/\lambda$) and the oscillation angle to match the $x$-direction ($\alpha = 0^\circ$). We apply this procedure on both the $\partial_x E_{\xNV}$ and $\partial_x E_{\yNV}$ images and average the results.

{\bf Surface charge density reconstruction}

To reverse propagate our $E$-field measurements into a surface charge density (Fig.~\ref{fig2}g), we first treat Coulomb's law for a two-dimensional surface charge density $\sigma (x,y)$ as a convolution integral in Fourier space~\cite{beardsley89}. With the transfer function $G(K, z) = e^{-Kz}/(2 \ep K)$, where $\ep$ is the vacuum permittivity, the $E$-field components from a single surface charge density become
\begin{equation}
	\mathcal{F}\{{\vecE}^\mr{lab}\} = \mathcal{F}\{\sigma\}G(K, z) {\vecK} \ .
\end{equation}
For two surface charge densities of opposite polarity separated by a distance $t$, the Fourier transformed $E$-field is $\mathcal{F}\{{\vecE}^\mr{total}\} = \mathcal{F}\{{\vecE}^\mr{lab}\} (1 - e^{-Kt})$, where the $-e^{-Kt}$ term comes from the bottom surface. The surface charge density can be computed for each of the three laboratory frame vector components and from our NV electrometry measurements the surface charge density is
\begin{multline}
	\mathcal{F}\{\sigma\} = \frac{2 \ep K}{3 \xosc k_r} \frac{e^{Kz}}{1 - e^{-Kt}} \cdot \\
	\left( \frac{\mathcal{F}\{\partial_r E_x^\mr{lab} \}}{k_x} + \frac{\mathcal{F}\{ \partial_r E_y^\mr{lab} \} }{k_y} + \frac{i\mathcal{F}\{ \partial_r E_z^\mr{lab} \}}{K}\right) \ ,
	\label{eq:sigmaft}
\end{multline}
where the three vector components have been averaged. We apply an additional window function, $W(\lambda_1, \lambda_2, \alpha)$, to Eq.~(\ref{eq:sigmaft}) that cuts off both low ($\lambda_1 = 30z$) and high ($\lambda_2 = z$) frequency components, and a line filter ($\alpha = 0^\circ$) to remove amplified noise from the deconvolution process.

{\bf Electric field gradiometry line scan fitting}
 
Fitting the line scans in Fig.~\ref{fig3}e is accomplished by first determining a simplified form for different surface charge models (Supplementary Section 2). For a monopole domain wall located at $x=x_i$ and propagating along $y$, the equations
\begin{equation}
\begin{split}
	\partial_x E_x^\mr{lab} = \frac{-\sigma}{\pi \ep} \frac{x-x_i}{(x-x_i)^2 + z^2} \ , \\
	\partial_x E_z^\mr{lab} = \frac{-\sigma}{\pi \ep} \frac{z}{(x-x_i)^2 + z^2} \ ,
	\label{eq:mono}
\end{split}
\end{equation}
are used. For the equivalent dipole domain wall the equations
\begin{equation}
	\begin{split}
	\partial_x E_x^\mr{lab} = \frac{\sigma d }{\pi \ep} \frac{2(x-x_i)z}{((x-x_i)^2 + z^2)^2} \ , \\
	\partial_x E_z^\mr{lab} = \frac{\sigma d}{\pi \ep} \frac{z^2 - (x-x_i)^2}{((x-x_i)^2 + z^2)^2} \ , 
	\label{eq:dipo}
\end{split}
\end{equation}
are used. In both types of domain walls $\partial_x E_y^\mr{lab} = 0$.
Here, the stand-off $z$, surface charge density $\sigma$ (or surface dipole density $\sigma d$), domain wall locations ($x_1$, and $x_2$), and sample thickness $t$ (for the bottom layer) are used to create the different surface charge models. 
Next, the gradient components are projected onto the NV center's $x$- and $y$-unit vectors using $\partial_x E_{x_\mr{NV}} = \partial_x {\vecE}^\mr{lab} \cdot \hat{x}_\mr{NV}$ and $\partial_x E_{y_\mr{NV}} = \partial_x {\vecE}^\mr{lab} \cdot \hat{y}_\mr{NV}$. The unit vectors depend on the polar and azimuthual angles $\thetaNV$ and $\phiNV$. Then, the $E$-field angle is computed as $\phidE = \arctan(\partial_x E_{y_\mr{NV}}/\partial_x E_{x_\mr{NV}})$.
Finally, the measured signal is modeled by $E_\mr{meas} = \xosc \sqrt{\left(\partial_x E_{x_\mr{NV}} \right)^2 + \left(\partial_x E_{y_\mr{NV}} \right)^2 } \cos (2 \phiB + \phidE)$.
During the fitting process, the NV angles, magnetic bias field angle, sample thickness, stand-off distance and oscillation amplitude are kept constant, having been measured or determined previously. The surface charge density (or surface dipole density) and domain wall locations are fitted and the best model is determined by the shape, quality and polarity of the fit.

{\bf Estimation of sensitivity}

We estimate the sensitivity with two methods, first by error propagating the measurements counts used in the quantum phase computation, and second by taking line-by-line difference from two consecutive line scans and computing the standard deviation of the resulting trace~\cite{palm22}. As shown in Supplemental Section 3, our best sensitivities are achieved by using multi-pulse sequences with quantum phase accumulation across multiple oscillation periods~\cite{huxter22}. The error propagated sensitivity is $0.24\unit{kV\,cm^{-1}\,Hz^{-1/2}}$ and the line-by-line sensitivity is $0.29\unit{kV\,cm^{-1}\,Hz^{-1/2}}$.

%%%%%%%%%%%%% References
%\bibliography{library}
% for the arXiv
%apsrev4-2.bst 2019-01-14 (MD) hand-edited version of apsrev4-1.bst
%Control: key (0)
%Control: author (8) initials jnrlst
%Control: editor formatted (1) identically to author
%Control: production of article title (0) allowed
%Control: page (0) single
%Control: year (1) truncated
%Control: production of eprint (0) enabled
%

\end{document}